\documentclass[preprint,aps]{revtex4}

\usepackage{graphicx}% Include figure files
\usepackage{dcolumn}% Align table columns on decimal point
\usepackage{bm}% bold math
\usepackage{epsfig}
\usepackage{subfigure}

\def\be{\begin{eqnarray}}\def\ba{\begin{eqnarray}}
\def\ee{\end{eqnarray}}\def\ea{\end{eqnarray}}

\def\bq{\begin{equation}}
\def\eq{\end{equation}}
\def\no{\nonumber \\}

\begin{document}

\title{ The effect of gluon condensate on holographic heavy quark potential }

\author{Youngman Kim}
 \email{ykim@apctp.org}
 \affiliation{Asia Pacific Center for Theoretical Physics
and Department of Physics, Pohang University of Science and Technology,
Pohang, Gyeongbuk 790-784, Korea}
 \affiliation{School of Physics, Korea Institute for Advanced Study, Seoul 130-722, Korea}

\author{Bum-Hoon Lee}
 \email{bhl@sogang.ac.kr}
 \affiliation{Department of Physics, Sogang University, Seoul 121-742, Korea }
 \affiliation{CQUeST, Sogang University, Seoul 121-742, Korea }

\author{Chanyong Park}
 \email{cyong21@sogang.ac.kr}
 \affiliation{CQUeST, Sogang University, Seoul 121-742, Korea }
 \affiliation{Division of Interdisciplinary Mathematics, National Institute
for Mathematical Sciences, Daejeon 305-340, Korea }

\author{Sang-Jin Sin}
 \email{sjsin@hanyang.ac.kr}
 \affiliation{Department of physics, BK21 Program Division, Hanyang University, Seoul 133-791, Korea}

\begin{abstract}
The gluon condensate is very sensitive to the QCD deconfinement transition since its value changes drastically
with the deconfinement transition.
We calculate the gluon condensate dependence of the heavy quark potential
in AdS/CFT to study how the property of the heavy quarkonium is affected by a relic of the deconfinement transition.
We observe that the heavy quark potential becomes deeper as the value of the gluon condensate decreases.
We interpret this as a dropping of the heavy quarkonium mass just above the deconfinement transition.
We finally argue that dropping gluon condensate and pure thermal effect are competing each other in
the physics of heavy quarkonium at high temperature.
\end{abstract}

\maketitle

\section{Introduction}

The heavy quarkonium  is a good object to study the nonperurbative nature of QCD, see ~\cite{hQnR} for a review.
At zero temperature, for example, the charmonium spectrum reveals important information about
confinement and inter-quark potentials in QCD\cite{potential}.
The heavy quarkonium is one of main probes that provides us with information
about thermal properties of the quark-gluon plasma (QGP), see, for instance,~\cite{MS86, Satz07}.
Lattice calculations indicate  that the charmonium
states will remain bound up to about 1.6 to 2
times the critical temperature $T_c$ of the deconfinement transition\cite{AH04,Datta03}.
Recent studies based on a QCD sum rule approach  have claimed that the change in the
properties of heavy quarkonia around  $T_c$ could {\it effectively} be an order
parameter of the deconfinement transition~\cite{ML}. The claim is based on the following observation.
At $T_c$, the energy density $\epsilon$ and pressure $P$ of the QCD system increase drastically, which
closely related to a drop of the thermal gluon condensate  $ G^2(T)$ since
\ba
 G^2(T) \approx  G^2(0)-(\epsilon-3P)\, .
\ea
 For instance, lattice results on the gluon condensate at finite
temperature~\cite{LgC} indicate that the value of the gluon condensate
shows a drastic change around $T_c$ regardless of the number of quark flavor.
 The change in the gluon condensate leads to
a dropping heavy quarkonium mass around $T_c$ when
 we ignore shift in the width of the quarkonium~\cite{ML}.
 Note that in \cite{ML}, the mass drops significantly at or just above $T_c$.
 {}From these studies~\cite{ML, LgC}, we see that the gluon condensate is very sensitive to the
 deconfinement transition of QCD and could serve as a messenger for deconfinement transition,
 which could be observed through the heavy quarkonium in
 relativistic heavy ion collisions (RHIC).
 We note here that dropping heavy quarkonium mass at $T_c$
 has previously observed in a study based on a AdS/QCD model.
 In the study~\cite{KLL} based on the soft wall model~\cite{KKSS}, the dropping is due to a Hawking-Page transition, and so
 deconfinement transition, which is qualitatively consistent with QCD sum rule results.
 We remark here that although the deconfinement transition transition is driven by the temperature, the drastic change in the value of
 the gluon condensate is mostly due to the deconfinement transition itself since the value of the temperature does not change much
 near the transition point.

 In this work we study the effect of the gluon condensate on
a heavy-quark potential in AdS/CFT~\cite{HQpAdSCFT}.
 To this end,
we adopt a deformed AdS background with back-reaction due to the gluon condensate~\cite{gcM, gcMT, bak, KLPS}. For a deformed AdS with a non-trivial dilaton potential, we refer to, {\it e.g.},
\cite{GG08}.
We observe that as the gluon condensate decreases in deconfined phase, the potential becomes deeper.
 Since the mass of a heavy quarkonium $m_{QQ}$ is roughly $m_{ QQ}\approx 2m_Q +V_{QQ}+ {\rm kinetic ~energy}$,
 we could associate the deepening of the potential with dropping of the heavy quarkonium mass.  To see this more clearly,
 we solve Schr\"{o}dinger equation with our heavy quark potential in the framework of potential model approach.
Our observation could
 be interpreted as the dropping of the heavy quarkonium mass right after the deconfinement transition, which may be a
reminiscence  of the deconfinement transition since the value of the gluon condensate decreases much with the deconfinement transition.

\section{Holographic heavy quark potential}

\begin{figure}
\begin{center}
{\includegraphics[angle=0,
width=0.5\textwidth]{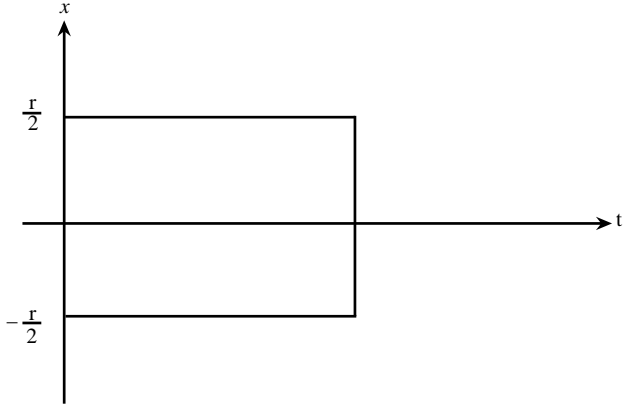}}
\caption{ A Wilson loop.}\label{WilsonL}
\end{center}
\end{figure}

Through the AdS/CFT correspondence, we can easily evaluate
the potential between heavy quark and anti-quark with a given background metric~\cite{HQpAdSCFT}.
 To obtain the heavy quark potential,
 we consider a Wilson loop living on the boundary of the five-dimensional AdS space
where the heavy quark and anti-quark are set at $x=r/2$ and $x=-r/2$, respectively, see Fig.~\ref{WilsonL}.

%%%
\begin{figure}
\begin{center}
\includegraphics{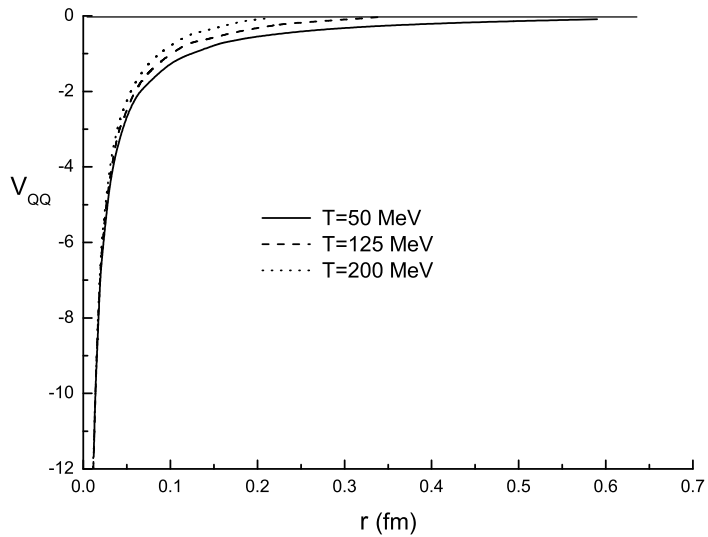}
\caption{ The heavy quark potential on the AdS black hole. Here $V_{QQ}\equiv\pi\alpha^\prime E_{\rm AdSBH}^R $.}\label{HQp:AdSBH}
\end{center}
\end{figure}

%%%
As a warm-up, we first work with the AdS black hole background in the Fefferman-Graham form.
The background is given by, in Euclidean,
\ba
ds^2=\frac{1}{z^2} \left( \frac{(1-az^4)^2}{1+az^4}dt^2 +(1+az^4)d\vec x^2 +dz^2 \right)\, ,\label{AdSBH}
\ea
where $a=(\pi T)^4/4$.
We calculate the heavy-quark potential following~\cite{HQpAdSCFT}.
The Nambu-Goto action for the open string connecting the boundary quark and the anti-quark
is given by
\be
S_{\rm NG} &=&\frac{1}{2\pi\alpha}\int d^2\sigma\sqrt{{\rm det}
G_{nm}\partial_\alpha X^n \partial_\beta X^m} . \label{S1}
\ee
After the gauge fixing,
\be \label{gf}
\sigma^0=t, \quad \sigma^1=x  \quad {\rm and} ~~ z=z(x) ,
\ee
the Nambu-Goto action on the background metric (\ref{AdSBH}) becomes
\be
S_{\rm AdSBH} &=& \frac{1}{2\pi\alpha} \int_{-T/2}^{T/2} dt \int_{-r/2}^{r/2}
dx \frac{1}{z^2} (1-az^4)\sqrt{\frac{1+az^4+{z^\prime}^2}{1+az^4}} \, ,
\ee
where $^{\prime}$ means the derivative with respect to $x$, and $T$ is the
time interval.
Considering the variable $x$ as a time variable, the
Hamiltonian of the system, which should be conserved, reads
\ba
\frac{1}{z^2}(1-az^8)\frac{1}{\sqrt{(1+az^4)(1+az^4+{z^\prime}^2)}}=const
=\frac{1}{z_0^2} (1-az_0^4)\, ,
\ea
where $z_0=z|_{x=0}$ with $z^\prime|_{x=0}=0$.
{}From this equation,
 we find the relation between inter-quark distance $r$ and $z_0$
\ba
r=2\int_0^{z_0} dz \frac{z^2}{z_0^2}\frac{1-az_0^4}{1-az^4}\frac{1}{\sqrt{1+az^4}}\frac{1}{\sqrt{1-
\frac{z^4(1-az_0^4)^2}{z_0^4 (1-az^4)^2}}}\, .
\ea
The regularized energy is
\ba
E_{\rm AdSBH}^R\equiv \frac{S_{\rm AdSBH}}{T} =\frac{1}{\pi\alpha^\prime}\int_0^{z_0} dz \frac{1-az^4}{z^2\sqrt{1+az^4}}
\frac{1}{\sqrt{1-\frac{z^4(1-az_0^4)^2}{z_0^4 (1-az^4)^2}}}
-\frac{1}{\pi\alpha^\prime}\int_0^{z_h} dz\frac{1}{z^2} \frac{1-az^4}{\sqrt{1+az^4}}\label{EAdSBH}
\, ,
\ea
where $z_h=1/a^{1/4}$, and the second term is corresponding to the masses of two free heavy quarks
at finite temperature~\cite{HQpAdSCFT}. So Eq.~(\ref{EAdSBH}) is nothing but the energy, equivalently potential,
 of heavy quark pair minus two free quark masses.
 Our results are shown in Fig.~\ref{HQp:AdSBH}. Note that in the figure we plot the potential given in Eq.~(\ref{EAdSBH}) only
 up to $r^\star$,  where the potential becomes zero, and for $r>r^\star$ the potential in Eq.~(\ref{EAdSBH}) becomes positive.
 The dissociation of heavy quarkonia happens when the potential becomes zero.
As expected, at low temperature the heavy quarkonia are hard to dissociate since $r^\star$ is rather big, for instance
$r^\star= 0.93~{\rm fm}$ at  T=50~{\rm MeV}.
 As the temperature increases, $r^\star$ decreases, and so the dissociation
is more likely to happen, for example, $r^\star= 0.24~{\rm fm}$ at  T=200~{\rm MeV}. We remark here that
to be more realistic, we should be able to distinguish the dissociation temperature of charmonium from that
of bottomonium. For this we may need to introduce an energy scale such as an infrared cutoff
other than the temperature to fix the scale of the system at hand.

\section{Gluon condensation and heavy quark potential}
To study the effect of the gluon condensation on the heavy quark potential, we consider
the 5D gravity action in Euclidean with a dilaton coupled
\be
S = \frac{1}{2 \kappa^2} \int d^5 x \sqrt{g} \left ( -{\cal R} - \frac{12}{R^2} + \frac{1}{2} \partial_M \phi \partial^M \phi  \right ) \, ,
\ee
where $\kappa^2$ is the 5D Newton constant, and $R$ is the AdS curvature.
By solving the coupled equations, dilaton equation of motion and the Einstein equation, with a suitable ansatz,
we obtain two relevant backgrounds.
The dilaton-wall solution is given by~\cite{gcM}
\ba
ds^2&=&\frac{R^2}{z^2}\left ( \sqrt{1-c^2z^8} (d\vec x^2 +dt^2) +dz^2 \right )\, ,\no
\phi(z)&=&\sqrt{\frac{3}{2}}\log \left ( \frac{1+cz^4}{1-cz^4} \right ) +\phi_0
\ea
where $\phi_0$ is a constant and $c=1/z_c^4$.
Another one is the dilaton black hole solution \cite{KLPS,bak},
\ba
ds^2=\frac{1}{z^2}\left( Ad\vec x^2 + Bdt^2 +dz^2\right )\, ,
\ea
where
\ba
A&=&(1+fz^4)^{(f+a)/2f} (1-fz^4)^{(f-a)/2f}\no
B&=&(1+fz^4)^{(f-3a)/2f} (1-fz^4)^{(f+3a)/2f}\,\no
&&f^2=a^2+c^2\, ,
\ea
and the corresponding dilaton profile is given by
\ba
\phi(z) &=& \phi_0 +
    \frac{c}{f}\sqrt{\frac{3}{2}} \log{ \left ( \frac{1 + f  {z^4} }{1 - f  {z^4} } \right )}.
\ea
Here $a$ is a temperature if $c=0$.
In both backgrounds, by expanding the dilaton profile near the boundary $z=0$, we obtain
 \ba
 \phi(z)=\phi_0+\sqrt{6}c z^4+\dots\, ,
 \ea
 and
 we can see that $c$ is nothing but the gluon condensation up to a constant by the AdS/CFT dictionary.
 In this work, we take $\phi_0=0$.

 As shown in \cite{KLPS}, there is a Hawking-Page transition between the thermal dilaton-wall solution and
 the dilaton black hole solution at some critical value of $a$.
 Thus, the dilaton-wall background is for confined phase, and the dilaton black hole describes deconfined phase.

\subsection{Dilaton wall background}

First, we consider the Euclidean dilaton-wall background~\cite{gcM}
\ba
ds^2=\frac{R^2}{z^2}\left ( \sqrt{1-c^2z^8} (d\vec x^2 +dt^2) +dz^2 \right )\label{dAdS}\, .
\ea
 {}From now on we will take $R=1$ for simplicity.
With the gauge fixing
\be
\sigma^0=t, \quad \sigma^1=x  \quad {\rm and} ~~ z=z(x) ,
\ee
the Nambu-Goto action on the background metric (\ref{dAdS}) becomes
\be
S_{\rm DAdS} &=& \frac{1}{2\pi\alpha} \int_{-T/2}^{T/2} dt \int_{-r/2}^{r/2}
dx e^{\frac{\phi}{2}}\frac{1}{z^2}\sqrt{1-c^2z^8+ \sqrt{1-c^2z^8} {z^\prime}^2}\, .
\ee
The first integral, Hamiltonian, is then
\ba \label{ham1}
H = -\frac{1}{2\pi\alpha} \frac{1-c^2z^8}{z^2\sqrt{1-c^2z^8+\sqrt{1-c^2z^8}{z^\prime}^2}} .
\ea
Due to the conserved Hamiltonian, after
comparing the Hamiltonian in (\ref{ham1}) with that evaluated at $z=z_0$
\be
H = -\frac{1}{2\pi\alpha} \frac{\sqrt{1-c^2 z_0^8}}{z_0^2} ,
\ee
we can easily find the integral relation between $r$ and $z_0$
\ba
r=2  \int_0^{z_0} dz \frac{z^2}{z_0^2}
\frac{\sqrt{1-c^2 z_0^8} }{(1-c^2 z^8)^{1/4} \sqrt{\frac{\delta^2}{\delta_0^2}(1 - c^2 z^8)+ c^2 z_0^4 z^4 - \frac{z^4}{z_0^4}
}}\, ,
\ea
where
\ba
\delta=\left( \frac{1+cz^4}{1-cz^4} \right)^{\sqrt{3\over 8}}, \,~ ~\delta_0=\left( \frac{1+cz_0^4}{1-cz_0^4} \right)^{\sqrt{3\over 8}}\, .
\ea
The value of $c$ ($=1/z_c^4$) could be determined as follows.
As in the work of  Csaki and Reece~\cite{gcM},  the lightest glueball mass calculated on the dilaton wall background
is $\sim6.61/z_c$ which is compared with the lattice value
$\sim 1.73 ~{\rm GeV}$ to fix the value of $1/z_c$: $c\sim (0.26~{\rm GeV})^4$.
Another way to fix the value of $z_c$ is to fit a heavy quarkonium mass.
In \cite{KLL}, $z_c$ is fixed by the lowest $c\bar c$ mass, $3.096~{\rm GeV}$, and it is
given by $1/z_c=1.29~{\rm GeV}$.
%%%
\begin{figure}
\begin{center}
\subfigure[] {\includegraphics[angle=0,
width=0.45\textwidth]{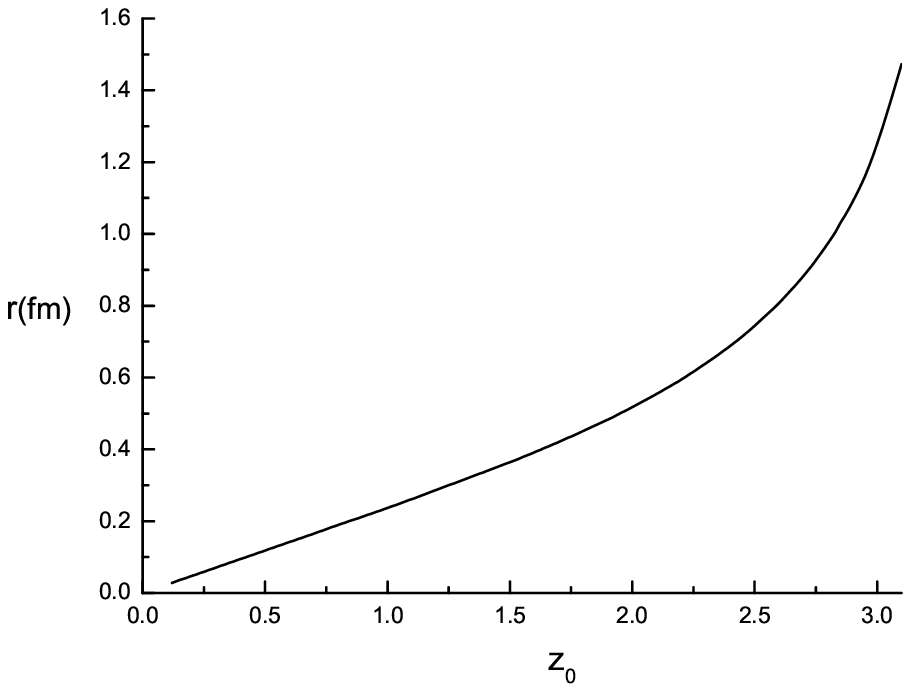} \label{cvr}}
\subfigure[] {\includegraphics[angle=0,
width=0.45\textwidth]{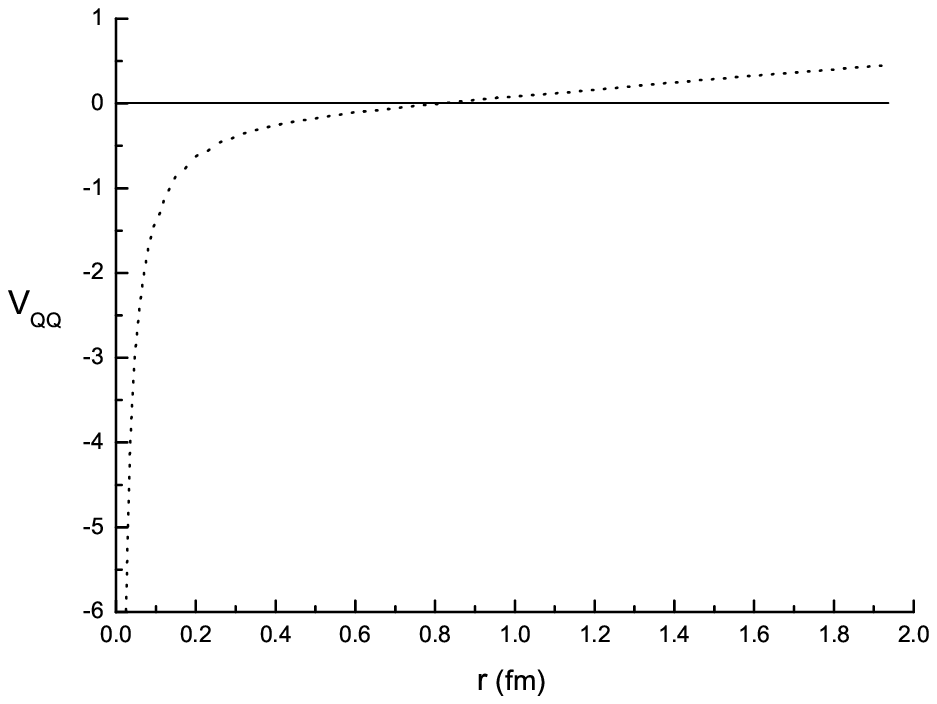} \label{rvV}}
\caption{ (a) Inter-quark distance $r$ as a function of $z_0$ with $1/z_c=0.26~{\rm GeV}$, where,
and (b) the corresponding heavy quark potential. }\label{DAdSr}
\end{center}
\end{figure}
%%%
The regularized energy is given by
\ba
E^R_{\rm DAdS} &=& \frac{1}{\pi\alpha^\prime}
\int_0^{z_0} dz\frac{\delta^2}{\delta_0} \frac{(1-c^2 z^8)^{1/4}}{z^2}
\frac{(1-c^2 z^8)^{1/2}}{\sqrt{ \frac{\delta^2}{\delta_0^2}(1 - c^2 z^8) + c^2 z_0^4 z^4
- \frac{z^4}{z_0^4} }} \no
&& -  \frac{1}{\pi\alpha^\prime} \int_0^{z_c} dz \delta\frac{(1-c^2 z^8)^{1/4}}{z^2} ,\label{Econfined}
\ea
where $z_c$ behaves as an IR cutoff, so $z$ and $z_0$ should be defined  in
the region smaller than $z_c$.
We show $r$ as a function of $z_0$ and plot the potential in Fig.~\ref{DAdSr}.

\subsection{Dilaton black hole background}\label{ssDBH}
Now, we move on to the dilaton black hole background,
\ba
ds^2=\frac{1}{z^2}\left( Ad\vec x^2 + Bdt^2 +dz^2\right )\, ,
\ea
where
\ba
A&=&(1+fz^4)^{(f+a)/2f} (1-fz^4)^{(f-a)/2f}\no
B&=&(1+fz^4)^{(f-3a)/2f} (1-fz^4)^{(f+3a)/2f}\,\no
&&f^2=a^2+c^2\, .
\ea
This dilaton black hole solution becomes the AdS black hole solution
when $c=0$, and it reduces to the dilaton wall background with $a=0$.
According to the Hawking-Page analysis done in \cite{KLPS}, this metric is for deconfined phase.
For discussion on the phases associated with
the dilaton black hole background in other context, we refer to \cite{bak}.
As discussed,  $c$ is  the gluon condensation.
Although the metric does not allow to define the Hawking temperature~\cite{gcMT, KLPS} by requiring absence of
conical singularity as long as $c\neq 0$,
we could associate $a$ with a temperature~\cite{KLPS}.
For more details about thermodynamics on the dilaton black hole
background, we refer to \cite{KLPS}.
On the dilaton black hole background, the Nambu-Goto action is given by
\be
S_{\rm DAdSBH}&=&\frac{T}{2\pi\alpha^\prime} \int_{-r/2}^{r/2} dx e^{\frac{\phi}{2}} \frac{1}{z^2} \sqrt{AB
+ B {z^{\prime}}^2} \, ,
\ee
where we used the same gauge fixing in (\ref{gf}). Then, the conserved Hamiltonian gives
rise to the connection between $z$ and $x$
\ba
z^\prime&=& \frac{1}{z^2\sqrt{B_0A_0}}\sqrt{(z_0^4BA-z^4B_0A_0) A}\, ,
\ea
where $A_0=A_{z=z_0}$ and $B_0=B_{z=z_0}$.
From this, we obtain a relation between $r$  and  $z_0$.
\be
r&=&2 \int_0^{z_0} dz z^2\sqrt{B_0A_0}\frac{1}{\sqrt{(z_0^4BA\frac{\gamma^2}{\gamma_0^2}-z^4B_0A_0) A}}\, ,
\ee
where
\ba
\gamma=\left( \frac{A}{B} \right)^{\sqrt{\frac{3}{2}}\frac{c}{4a}},\, \gamma_0=\left( \frac{A_0}{B_0} \right)^{\sqrt{\frac{3}{2}}\frac{c}{4a}}\, .
\ea
Hence, after using the similar regularization method in the previous section, we
finally obtain the regularized energy on the dilaton black hole background
\be
E^R_{\rm DAdSBH}&=& \frac{1}{\pi\alpha^\prime} \int_0^{z_0} dz
\frac{\gamma^2}{\gamma_0}\frac{\sqrt{B}}{z^2}
\frac{\sqrt{AB}}{\sqrt{ \frac{\gamma^2}{\gamma_0^2} AB-A_0 B_0 \frac{z^4}{z_0^4}}}
- \frac{1}{\pi\alpha^\prime} \int_0^{z_f} dz \gamma\frac{\sqrt{B}}{z^2}\, ,\label{EDAdSBH}
\ee
where $z_f$ is the position of the IR cutoff defined by $z_f = f^{-1/4}$.

%%%
\begin{figure}[!ht]
\begin{center}
\includegraphics{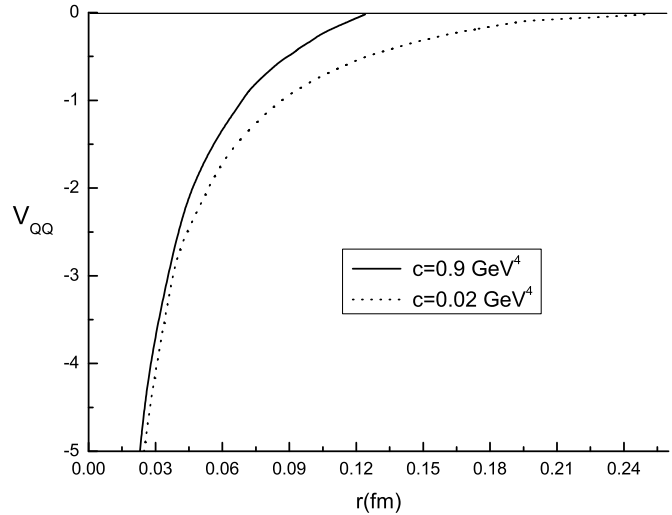}
\caption{\small  The gluon condensate $c$ dependence  of the heavy quark potential with fixed $a$. Here $a=(\pi T)^4/4$,
and we take $T=200 ~{\rm MeV}$. }\label{Fig:DAdSBH}
\end{center}
\end{figure}
%%%
In Fig.~\ref{Fig:DAdSBH}, we plot the heavy quark potential with a fixed $a$ for a few  values of $c$.
{}From Fig.~\ref{Fig:DAdSBH}, we can see that  the
heavy quark potential becomes deeper as $c$ decreases,
implying the value of the gluon condensate drops in deconfined phase, the mass of heavy quarkonium
decreases. For the heavy quarkonium, this is so since the mass of it is roughly equal to the masses of two
heavy quarks plus the potential.
 Noting that the value of the gluon condensate suddenly drops near $T_c$ of QCD deconfinement transition~\cite{LgC, ML},
we may conclude that our study predicts dropping quarkonium mass just after the deconfinement transition.

\begin{figure}
\begin{center}
\includegraphics{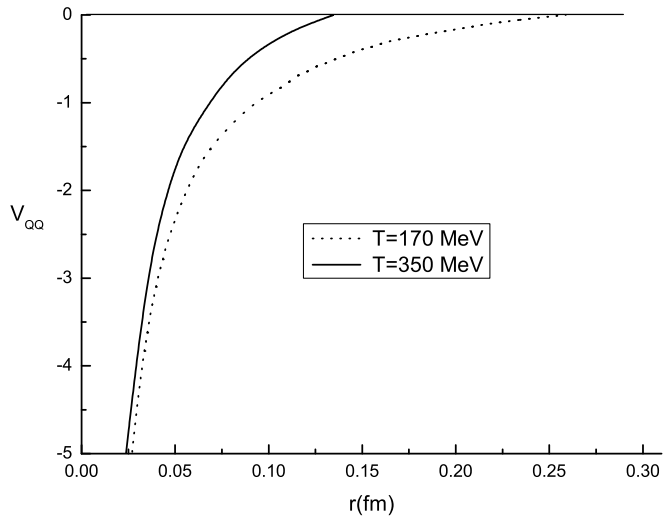}
\caption{ The temperature dependence of the heavy quark potential with a fixed gluon condensate $c$. Here we take $c=0.02~ {\rm GeV}^4$.
}\label{c02}
\end{center}
\end{figure}

Finally,  we study the $a$ dependence of the potential with fixed $c$.
 Note that $a$ is related to a temperature, $a\sim T^4$,~\cite{KLPS}.
As shown in Fig.~\ref{c02}, the potential becomes shallow as $a$ (temperature) increases. This means that
the mass of heavy quarkonium increases as we increases the temperature, which is consistent with
the result in~\cite{KLL}. As expected the value of $r^\star$
becomes smaller with increasing temperature, and so easy dissociation of heavy quarkonium at high temperature.
%%%
%%%

\subsection{Heavy quarkonium and gluon condensate}
Using the heavy quark potential obtained in section \ref{ssDBH}, we estimate the mass of heavy
quarkonium by solving the Schr\"{o}dinger equation, following potential model approach.
In potential models, it is assumed that the interaction between two heavy quarks in a heavy quarkonium
could be described by a heavy quark potential. Since the mass of heavy quarks are much larger than a typical scale
of QCD, $m_{\rm c,b}>> \Lambda_{\rm QCD}$, the heavy quark system is generically  non-relativistic.

Our aim in this section is not to give exact numbers for the masses, but to see the role of  gluon condensate in the heavy-quark system.
To this end we follow an approximate method given in \cite{IS} and consider only the leading order in \cite{IS} for simplicity.
Here we briefly summarize the method and refer to \cite{IS} and references there in for details.
To use the method, potentials should satisfy the following conditions:
\ba
&&V(r)=-Ar^{-\alpha }+\kappa r^{\beta }+V_{0},~\alpha ,\beta >0\, ,\no
&& V^{\prime }(r)>0\text{ \ and }V^{\prime \prime }(r)\leq 0\,
\ea
where $V_{0}$ is a constant, and $A$ and $\kappa $ are positive constants.
We confirmed that our potential in section \ref{ssDBH} satisfies these requirements.
Then the binding energy is given by
\be
E_{n,l}\simeq V(r_{0})+\frac{1}{2}r_{0}V^{\prime }(r_{0}) \, ,\label{BEf}
\ee
where higher orders are neglected for simplicity.
$r_0$ is determined by
\be
1+2l+(2n_{r}+1)\left[ 3+\frac{r_{0}V^{\prime \prime }(r_{0})}{V^{\prime
}(r_{0})}\right] ^{1/2}=\left[ 8\mu r_{0}^{3}V^{\prime }(r_{0})\right]
^{1/2},\text{ }n_{r},l=0,1,2,3,\cdots \, .
\ee
Finally, we arrive at the mass of bound state:
\be
M_{\bar QQ}=2m_Q+2 E_{n,l}\, .
\ee
Now, as examples, we consider charm and bottom quarks.
%%%%%%%%%%%%%%%%%%%%%%%%%%%%%%%%%%%%%%%%%%%%%%%%%%%%%%%%%%%%%%%%%%%%%%%%%%%%
%\vspace{1cm}
\begin{center}
\begin{tabular}{|c||c|c|}
\hline
c (${\rm GeV}^4$)&  $r_0 ~({\rm GeV}^{-1})$ & binding energy ({\rm GeV}) \\
\hline  \hline
0.001&  $0.927$ & 0.208  \\
\hline
0.007  &  $0.937$ & 0.217  \\
\hline
0.02&  $0.982$ & 0.248  \\
\hline
\end{tabular}
\end{center}
%\vspace{0.3cm}
\noindent Table 1. Charm-quark system and gluon condensate. Here the charm quark mass is  $1.396 ~{\rm GeV}$
and $T=200~{\rm MeV}$. \\
\vskip 0.3cm
%%%%%%%%%%%%%%%%%%%%%%%%%%%%%%%%%%%%%%%%%%%%%%%%%%%%%%%%%%%%%%%%%%%%%%%%%%%%
In Table 1, we consider the charm-quark system. Since the binding energy is positive, we have no bound state out of charm quarks.
Here we don't claim that there is no charmonium in QGP since we adopted a very crude approximation.
What is interesting here is that smallness of gluon condensate supports the existence of charmonium since
the positive binding energy decreases with decreasing gluon condensate $c$.
This is, as it should be, consistent with our observation made in the previous section: decreasing $c$ means the deepening
in the potential.
But, we have no clear physical interpretation on this observation.

%%%%%%%%%%%%%%%%%%%%%%%%%%%%%%%%%%%%%%%%%%%%%%%%%%%%%%%%%%%%%%%%%%%%%%%%%%%%
%\vspace{1cm}
\begin{center}
\begin{tabular}{|c||c|c|c|c|c|c|}
\hline
c (${\rm GeV}^4$)& $r_0 ~({\rm GeV}^{-1}) $ & binding energy ({\rm GeV})
& meson mass  ({\rm GeV})\\
\hline  \hline
0.02  & $0.285$ & $-0.624 $ & $8.357 $ \\
\hline
0.2  & $0.291$ & $-0.465 $ & $8.675 $ \\
\hline
0.9  & $0.304$ & $-0.121 $ & $9.364 $ \\
\hline
\end{tabular}
\end{center}
%\vspace{0.3cm}
\noindent Table 2. Bottom-quark system and gluon condensate. Here the charm quark mass is  $4.803 ~{\rm GeV}$
and $T=200~{\rm MeV}$\\
\vskip 0.3cm
%%%%%%%%%%%%%%%%%%%%%%%%%%%%%%%%%%%%%%%%%%%%%%%%%%%%%%%%%%%%%%%%%%%%%%%%%%%%
In Table 3, we consider the bottom-quark system. As expected, the mass of bottomoinum decreases as the
gluon condensate is to be smaller.

\section{Summary and Discussion}
With an observation~\cite{LgC}  that the gluon condensate is an useful quantity to characterize the QCD deconfinement transition,
we calculate the holographic heavy quark potential on deformed AdS backgrounds with gluon condensate included.
We also solved the Schr\"{o}dinger equation approximately
to estimate the mass of heavy quarkonium, following the potential model approach.

Our analysis of the dilaton black hole metric with fixed $a$ reveals that the potential becomes deeper as the gluon condensate
 decreases in deconfined phase.
 We associate this with the dropping of the heavy quarkonium mass in deconfined phase.
 With the deconfinement transition, the value of the gluon condensate drops much, and therefore
 the dropping mass is most likely to occur just after the deconfinement transition.
  Finally, we study the $a$ dependence of the heavy quark potential and show that the potential becomes
  shallow as we increase $a$. If we interpret $a$ as a temperature~\cite{KLPS}, our result indicates that
  the mass of the heavy quarkonium increases with temperature in deconfined phase.

As we increase the temperature right after the deconfinement transition, the mass of a heavy quarkonium
increases with temperature, while it decreases with a decreasing gluon condensate.
  Just after the deconfinement transition,
   although the gluon condensate is not an order parameter of the deconfinement transition,
  its value does change dramatically with the deconfinement transition.
  Even though the deconfinement transition is driven by the temperature, the drastic drop in the value of
 the gluon condensate is from the deconfinement transition itself since the value of the temperature changes very little
 around the transition point.
  Therefore, right after the deconfinement transition, the effect of the gluon condensate should dominate over the
  temperature.

  In conclusion, based on the drastic drop of the gluon condensate right after the deconfinement transition observed in
  lattice QCD~\cite{LgC},
  our study predicts that the mass of heavy quarkonium just after the deconfinement transition
   decreases  with decreasing gluon condensate, which is consistent with \cite{ML, KLL}.
   As we increase the temperature further above $T_c$, gluon condensate reduces to increase the mass of heavy quarkonium,
   while increasing $T$ will make it heavier. Therefore, to reach a concrete conclusion on the mass of heavy quarkonium
   at high temperature, except very close to $T_c$, we have to perform detailed study on the competition
   of the two effects: effects of the gluon condensate and temperature (pure thermal).
   To this end, it is essential to calculate the temperature dependence of the gluon condensate in AdS/CFT.

\vskip 1cm
%{\bf acknowledgments}
\begin{acknowledgments}
YK would like to thank Su Houng Lee and Kenji Morita for useful discussions on charmonium
near the deconfinement transition.
This work was supported in part by
the Science Research Center Program of the Korea Science and Engineering Foundation through
the Center for Quantum Spacetime(CQUeST) of Sogang University with grant number R11 - 2005 - 021.
YK acknowledges the Max Planck Society(MPG) and the Korea Ministry of
Education, Science and Technology(MEST) for the support of the Independent
Junior Research Group at the Asia Pacific Center for Theoretical Physics
(APCTP).
C. Park was partially supported by the Korea
Research Council of Fundamental Science and Technology (KRCF).
\end{acknowledgments}

%%%%%%%%%%%%%%%%%%%%%%%%%%%%%%%%%%%%%%%%%%%%%%%%%%%%%%%%%%%%%%%%%%%%%%%%%%%%%%%%%%%%%%%%%%%%%%%%%%%%%%%
%                                                                                                     %
%    The Bibliography                                                                                 %
%                                                                                                     %
%%%%%%%%%%%%%%%%%%%%%%%%%%%%%%%%%%%%%%%%%%%%%%%%%%%%%%%%%%%%%%%%%%%%%%%%%%%%%%%%%%%%%%%%%%%%%%%%%%%%%%%

\end{document}